# Electronic structure contribution to hydrogen bonding interaction of a water dimer

Zhiyuan Zhang,[1,2] Wanrun Jiang,[1,2] Bo Wang,[1,2] and Zhigang Wang[1,2,a]

[1]*Institute of Atomic and Molecular Physics, Jilin University, Changchun 130012, People's Republic of China*

[2]*Jilin Provincial Key Laboratory of Applied Atomic and Molecular Spectroscopy (Jilin University), Changchun 130012, People's Republic of China*

Hydrogen bond (H-bond) covalency has recently been observed in ice and liquid water, while the penetrating molecular orbitals (MOs) in the H-bond region of most typical water dimer system, $(H_2O)_2$, have also been discovered. However, obtaining the quantitative contribution of these MOs to the H-bond interaction is still problematic. In this work, we introduced the orbital-resolved electron density projected integral (EDPI) along the H-bond to approach this problem. The calculations show that, surprisingly, the electronic occupied orbital (HOMO-4) of $(H_2O)_2$ accounts for about 40% of the electron density at the bond critical point. Moreover, the charge transfer analysis visualizes the electron accumulating effect of the orbital interaction within the H-bond between water molecules, supporting its covalent-like character. Our work expands the classical understanding of H-bond with specific contributions from certain MOs, and will also advance further research into such covalency and offer quantitative electronic structure insights into intermolecular systems.

The hydrogen bond (H-bond) is one of the most important interactions in the nature and has long been at the research frontier in many fields, including molecular physics, chemistry and life sciences[1-4]. Water dimer $(H_2O)_2$, as the most typical H-bond system, has been widely studied in both theoretical and experimental works[5-10]. A number of its properties, including its geometry, electronic parameters[11, 12], quantum tunneling[13] and O-H stretching vibration modes[14-16], have been discovered. There is also a research which has found anti-electrostatic effects in many ionic H-bond systems[17]. These properties are critical for further understanding the behavior of H-bonds. An experimental probe of the covalency of H-bonds in liquid water has been recently established through proton nuclear magnetic resonance[18]. Meanwhile, similar covalency has also been directly measured between H-bonded molecules, for example, in ice using X-ray scattering technology[19]. Our previous theoretical work revealed that the delocalized molecular orbitals (MOs) of (H$_2$O)$_2$, which are of lower energy than the highest occupied molecular orbitals (HOMO), penetrate the H-bond region, as a reflection of the covalent-like character[20]. However, the quantitative contribution from the covalency to H-bond is still not clear in the above studies. In this work, we implemented a procedure to separate the electron density distribution along the H-bond into specific MOs contributions based on the electron density projected integral (EDPI) of the delocalized orbitals, thus revealing their weights to the H-bond interaction from an electronic structure view point. Our results provide further understanding of the H-bond and indicate that delocalized MOs play important roles in H-bond systems.

---

[a] Electronic mail: wangzg@jlu.edu.cn

Since the MOs obtained from both Hartree–Fock and density functional theory (DFT) show similar morphologies and properties, the latter was adopted in this research. The calculations were performed using the Gaussian 09 program[21] with the PBE0 method[22-24] and aug-cc-pVQZ basis set [25, 26]. The related analyses of the EDPI and LDOS were aided by the Multiwfn 3.3.8 package[27]. To make the following descriptions easier to understand, the O atom of the acceptor water monomer was placed at the origin, while the H atom of the donor monomer on the H-bond was on the y-axis (see Fig. 1).

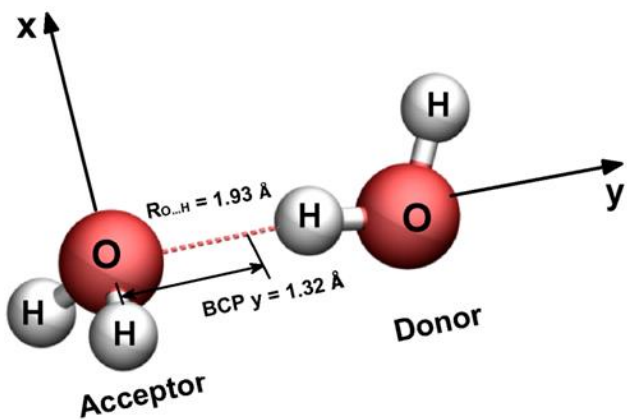

FIG. 1. Spatial position and coordinate system of (H$_2$O)$_2$. The O atom of the acceptor is located at the origin and the H atom of the H-bond is located at y = 1.93 Å on the axis. The bond critical point (BCP) of the H-bond is at y = 1.32 Å on the axis.

In order to obtain the spatial distribution properties of the MOs, we conducted a topological analysis[28] of the real-space electron densities of the $(H_2O)_2$ MOs. An extreme point can be found on the H-bond path when y = 1.32 Å, namely, the bond critical point (BCP) (see Fig. 2). The percentage for the local EDPI of a certain MO indicates its contribution among all of the MOs within the area. Most of the orbitals have electron densities localized near the O and H atoms, while in the intermolecular region, HOMO-2, and in particular HOMO-4, have stronger contributions compared to the other orbitals.

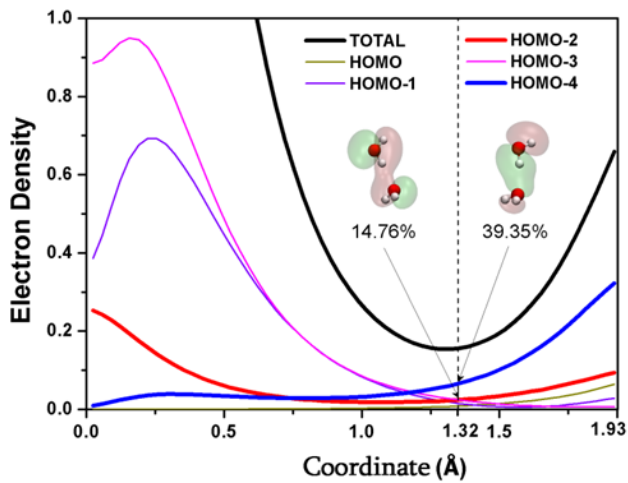

FIG. 2. Electron density integral of the MOs projected on the H-bond. The vertical axis represents the integrated electron density, and the horizontal axis indicates the y-axis on which the O atom of the acceptor is located at y = 0 and the H atom of the H-bond is located at y = 1.93 Å. The curve is obtained by integrating the electron density in the x and z directions onto the y direction, which can be written as $I(y) = \int_{-\infty}^{+\infty} \rho(x,y,z)\,dxdz$. The vertical dashed line represents the position of the BCP.

According to the EDPI, the percentages of the HOMO-2 and HOMO-4 electron density contributions to the total can be obtained. At the BCP of the H-bond, over half of the electron density comes from HOMO-2 and HOMO-4, in which HOMO-4 accounts for 39.35% of the total and has the largest contribution. However, the percentage of HOMO-2, beneath 15%, is much closer to the values of HOMO-1 and HOMO-3, less than HOMO-4.

In fact, corresponding quantitative contributions have been reported by using natural bond critical point (NBCP) method and natural bond orbital (NBO) decomposition, indicating large contributions from a lone pair NBO of O atom in the acceptor and a $\sigma_{OH}$ bond from the doner[29]. These ingenious methods expand the research means and are of great significance for studying the bonding properties in various systems, especially in those involving H-bonds. However, it is the contribution from the penetrating MOs based on the energy eigenvalue that we really considered about, which could be given by the EDPI method discussed above. Moreover, the EDPI method is also capable of giving the trend of electron density along a certain real-space direction, helpful for the correspoding spatial investigations.

To obtain a more comprehensive view of the covalent-like binding character in $(H_2O)_2$, we further extracted the electronic structure response to the orbital interaction from other interaction effects in the form of dimerization charge transfer separations. The total charge transfer shows electron collection behavior in the intermolecular region, suggesting a bonding-like effect that corresponds to conventional concepts, as shown in Fig. 3(a). The charge transfer of the orbital interaction also shows bonding-like electron accumulation between the O and H atoms of H-bond as expected (see Fig. 3(b)), and conversely, other effects (i.e. the electrostatic attraction and Pauli repulsion) actually suggest a decreasing trend in electrons, as shown in Fig. 3(c).

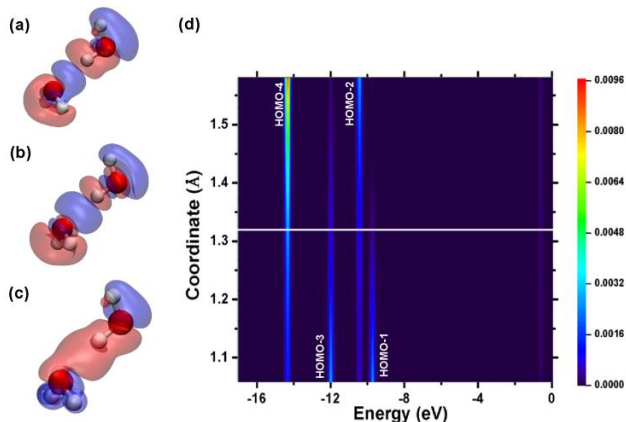

FIG. 3. Electronic density response to dimerization and the real-space local density of states (LDOS) of $(H_2O)_2$. (a) Total charge transfer from the two independent monomers to the complex. (b) Charge transfer of the orbital interaction in dimerization, obtained as the electron difference before and after the self-consistent field (SCF) is applied (SCF iteration on the well-constructed dimeric wavefunction). (c) Charge transfer from two single monomers to construct a dimeric wavefunction while forbidding consecutive orbital SCF iterations. (d) LDOS of the H-bond along the bond path. The horizontal axis indicates the energy of the states and the vertical axis represents the coordinates of the y-axis, while the color bar describes the density of each state. The white line shows the position of the BCP.

In order to further establish the electron distribution properties in the H-bond region, we studied the density of states (DOS) of $(H_2O)_2$. Here, focusing on the interaction properties, we adopted the local DOS (LDOS) along the H-bond, in which the different distributions of the LDOS for each MO over the coordinates are displayed. The LDOS indicates that HOMO-4 has a large contribution near the BCP of the H-bond and shows a relatively strong crossing character through the entire intermolecular region, and it also has a more delocalized feature compared with other MOs (see Fig. 3(d)). However, the other crossing MO, HOMO-2, makes a lesser contribution in this area.

In conclusion, our study clearly indicates that among these electronic occupied molecular orbitals, HOMO-4 is primarily responsible for the covalent-like characteristic of H-bond due to its strong contribution within the H-bond region, and the LDOS analysis also shows its delocalized and bonding-like behavior at the BCP. Meanwhile, the result of the charge transfer indicates that the dimerization electron accumulation in $(H_2O)_2$ mainly comes from the orbital interactions. As a consequence, it can be concluded that HOMO-4 plays the dominating role in the intermolecular H-bond electronic structure of $(H_2O)_2$. Although HOMO-2 morphologically presents a crossing character along the H-bond, its behavior is more like the localized HOMO-1 and HOMO-3, as shown by the LDOS. Therefore, penetrating MOs, which reflects orbital interaction, is probably the key to the covalent character of H-bond, like HOMO-4 in $(H_2O)_2$. However, although several works on the quantum effects of H-bond have been reported, including the electron orbital effect and nuclear vibration properties, more effort is still needed to develop a complete understanding. We hope that the results from this work will encourage further research in this field.

**ACKNOWLEDGMENTS**

We would like to acknowledge the National Natural Science Foundation of China (grant numbers 11374004 and 11674123) for the support. Z. W. also acknowledges the Fok Ying Tung Education Foundation (142001) and the High Performance Computing Center of Jilin University.